\documentclass[aps,preprintnumbers,eqsecnum,amsmath,amssymb,showpacs,nofootinbib]{revtex4}
\usepackage{graphicx}
\usepackage{dcolumn}
\usepackage{bm}
\usepackage{epsfig}

\begin{document}
\title{\boldmath Decay constants of the charmed vector mesons $D^*$ and $D^*_s$ from QCD sum rules}
\author{Wolfgang Lucha$^{a}$, Dmitri Melikhov$^{b,c}$, and Silvano Simula$^{d}$}
\affiliation{
$^a$HEPHY, Austrian Academy of Sciences, Nikolsdorfergasse 18, A-1050, Vienna, Austria\\
$^b$D.V.Skobeltsyn Institute of Nuclear Physics, M.V.Lomonosov Moscow State University, 119991, Moscow, Russia\\
$^c$Faculty of Physics, University of Vienna, Boltzmanngasse 5, A-1090 Vienna, Austria\\
$^d$INFN, Sezione di Roma III, Via della Vasca Navale 84, I-00146, Roma, Italy}
\date{\today}
\begin{abstract}
We present a sum-rule calculation of the decay constants of the
charmed vector mesons $D^*$ and $D^*_s$ from the two-point
correlator of vector currents. First, we show that the
perturbative expansion in terms of the pole mass exhibits no sign
of convergence whereas the reorganization of this expansion in
terms of the $\overline{\rm MS}$ mass leads to a distinct
hierarchy. Second, making use of the operator product expansion in
terms of the $\overline{\rm MS}$ mass, we determine the decay
constants of the $D^*$ and $D^*_s$ mesons with an emphasis on the
uncertainties in these theoretically predicted quantities related
both to the input QCD parameters and to the limited accuracy of
the method of sum rules. Our results are 
$f_{D^*}=(252.2 \pm 22.3_{\rm OPE}\pm 4_{\rm syst})\; {\rm MeV}$ and 
$f_{D_s^*}=(305.5 \pm 26.8_{\rm OPE}\pm 5_{\rm syst})\; {\rm MeV}$. For the ratios
of the vector-to-pseudoscalar decay constants we report
$f_{D^*}/f_D= 1.221\pm 0.080_{\rm OPE}\pm 0.008_{\rm syst}$ and
$f_{D^*_s}/f_{D_s}= 1.241\pm 0.057_{\rm OPE}\pm 0.007_{\rm syst}$.
\end{abstract}
\maketitle

\section{Introduction}
The extraction of the decay constants of ground-state vector
mesons within the method of QCD sum rules \cite{svz,aliev} is
based on the analysis of the two-point correlation function
\begin{eqnarray}
\label{1.1} \label{Pi_QCD} 
i \int d^4x\, e^{ipx}\langle
0|T\!\left(j_\mu(x)j^\dagger_\nu(0)\right)| 0\rangle=
\left(-g_{\mu\nu}+\frac{p_\mu p_\nu}{p^2}\right) \Pi(p^2)+\frac{p_\mu p_\nu}{p^2} \Pi_L(p^2)
\end{eqnarray}
of the vector heavy--light currents for a heavy quark $Q$ of mass
$m_Q$ and a light quark $q$ of mass $m$
\begin{eqnarray}
j_\mu(x)=\bar q(x) \gamma_\mu Q(x),
\end{eqnarray}
or, more precisely, on the Borel transform $\Pi(p^2)\to \Pi(\tau)$
of its transverse structure to Borel variable $\tau$.
Equating~$\Pi(\tau)$ as calculated within QCD and the expression
obtained by inserting a complete set of hadron states yields the
sum~rule
\begin{eqnarray}
\label{pitau} \Pi(\tau)=f^2_{V}M_V^2
e^{-M_V^2\tau}+\int\limits_{s_{\rm phys}}^{\infty}ds\, e^{-s \tau}
\rho_{\rm hadr}(s) = \int\limits^\infty_{(m_Q+m)^2}ds\,
e^{-s\tau}\rho_{\rm pert}(s,\mu) + \Pi_{\rm power}(\tau,\mu).
\end{eqnarray}
Here, $M_V$ is the mass, $f_V$ the decay constant, and
$\varepsilon_\mu(p)$ the polarization vector of the vector meson
$V$ under study:
\begin{eqnarray}
\label{decay_constant} \langle 0 |\bar q \gamma_\mu Q| V(p)\rangle
= f_{V} M_{V}\varepsilon_\mu(p).
\end{eqnarray}
For the correlator (\ref{1.1}), $s_{\rm phys}=(M_{P}+M_\pi)^2$ is
the physical continuum threshold, wherein $M_{P}$ denotes the mass
of the pseudoscalar meson containing $Q$. For large values of
$\tau$, the ground state dominates the correlator and thus its
properties may be calculated from the correlation function
(\ref{1.1}).

In perturbation theory, the correlation function is obtained as
expansion in powers of the strong coupling constant $\alpha_{\rm
s}(\mu)$. The best known three-loop perturbative spectral density
has been calculated in \cite{chetyrkin} in terms of the
\emph{pole\/} mass~of the heavy quark $Q$ (called $M$ here) and
for a massless second quark [$\alpha_{\rm s}(\mu)$ is the running
coupling constant in the~$\overline{\rm MS}$ scheme]:
\begin{eqnarray}
\label{rhopert}
\rho_{\rm pert}(s)=
\rho^{(0)}(s,M)+\frac{\alpha_{\rm s}(\mu)}{\pi}\rho^{(1)}(s,M)+
\left(\frac{\alpha_{\rm s}(\mu)}{\pi}\right)^2 \rho^{(2)}(s,M,\mu)+\cdots.
\end{eqnarray}
For two massive quarks, the two-loop spectral density in terms of
their pole masses was obtained in \cite{rubinstein}.

However, already for the case of the pseudoscalar correlator it
was found that the perturbative expansion in terms of the heavy-quark pole 
mass does not exhibit any sign of convergence; this
problem was cured by rearranging the perturbative expansion in
terms of the corresponding running $\overline{\rm MS}$ mass
\cite{jamin}. We show that precisely the same happens in the case
of the vector correlator (\ref{1.1}).

Another subtlety --- related to the truncation of the perturbative
expansion --- is the unphysical dependence of the obtained
ground-state parameters on the renormalization scale $\mu$: of
course, the full correlator (\ref{1.1}) does not depend on $\mu$;
however, both the perturbative expansion truncated at fixed order
in $\alpha_{\rm s}$ and the truncated power corrections $\Pi_{\rm
power}(\tau,\mu)$ depend on $\mu$. For the pseudoscalar-meson
decay constants, this dependence was found to be rather mild
\cite{lms_plb2011}. Unfortunately, as we shall demonstrate in this
analysis, for the vector-meson decay constants the $\mu$
dependence is rather pronounced; this leads to a larger
corresponding error in the decay constants of vector mesons
obtained from QCD sum rules.

Furthermore, the truncated operator product expansion (OPE) does
not allow one to calculate the correlator for sufficiently large
$\tau$, such that the continuum states give a sizable contribution
to $\Pi(\tau)$ in the corresponding $\tau$-range. In order to get
rid of the continuum contribution, the concept of duality is
invoked: Perturbative-QCD spectral density $\rho_{\rm pert}(s)$
and hadron spectral density $\rho_{\rm hadr}(s)$ resemble each
other at large values of $s$; thus, for sufficiently large values
of the parameter $\bar s$, (far) above the resonance region, one
arrives at the duality relation
\begin{eqnarray}
\label{duality1} \int\limits_{\bar s}^{\infty} ds\, e^{-s
\tau}\rho_{\rm hadr}(s) = \int\limits_{\bar s}^{\infty}ds\, e^{-s
\tau}\rho_{\rm pert}(s).
\end{eqnarray}
Now, in order to express the continuum contribution in terms of
the perturbative contribution, this relationship should be
extended down to the hadronic threshold $s_{\rm phys}$. However,
the spectral densities $\rho_{\rm pert}(s)$ and $\rho_{\rm
hadr}(s)$ are obviously different in the region near $s_{\rm
phys}$. Therefore, one can only expect to obtain a relation of the
form
\begin{eqnarray}
\label{duality}
\int\limits_{s_{\rm phys}}^{\infty} ds\, e^{-s
\tau} \rho_{\rm hadr}(s) = \int\limits_{s_{\rm
eff}(\tau)}^{\infty} ds\, e^{-s \tau} \rho_{\rm pert}(s),
\end{eqnarray}
where $s_{\rm eff}(\tau)$ is different from the physical threshold
$s_{\rm phys}$. Obviously, for the same reason which causes $s_{\rm eff}(\tau)\ne s_{\rm phys}$, $s_{\rm eff}(\tau)$ must be a
function of the parameter $\tau$ \cite{lms_1,lms_new}. By virtue of (\ref{duality}), we may rewrite the sum rule
(\ref{pitau}) as
\begin{eqnarray}
\label{sr} 
f_V^2 M_V^2 e^{-M_V^2\tau}= \int\limits^{s_{\rm
eff}(\tau)}_{(m_Q+m)^2} ds\, e^{-s\tau}\rho_{\rm pert}(s,\mu) +
\Pi_{\rm power}(\tau,\mu) \equiv \Pi_{\rm dual}(\tau,s_{\rm eff}(\tau)).
\end{eqnarray}
We refer to the right-hand side of this equation as the {\it dual correlator} and to the $\tau$-dependent 
effective threshold that corresponds to the true values of the ground-state parameters in the left-hand side of (\ref{sr}) 
as the {\it exact effective threshold}; by definition, the exact effective threshold makes Eq.~(\ref{sr}) an identity. 
One essential property of the exact effective threshold should be mentioned: Whereas the exact correlation function and its 
truncated OPE have very different energy dependences in the Minkowski space, 
after performing the Borel transform $\Pi(p^2)\to \Pi(\tau)$, the complicated energy dependence of the exact correlation function 
leads to only a weak $\tau$-dependence of the exact effective threshold. This feature opens the possibility to find realistic 
approximations to this exact $\tau$-dependent threshold and to obtain in this way reliable estimates for the 
bound-state parameters. 

Obviously, the exact effective threshold is unknown. Thus, the extraction of the decay constant requires, in addition 
to $\rho_{\rm pert}(s,\mu)$ and $\Pi_{\rm power}(\tau,\mu)$, as further input, a criterion for obtaining an approximation 
to the exact effective threshold. In \cite{lms_new} we developed the algorithm for fixing $s_{\rm eff}(\tau)$ which allows one 
to reliably extract the ground-state parameters on the basis of (i) an accurate OPE for the Green functions and 
(ii) the known value of the ground-state mass. 

We shall demonstrate that QCD sum rules armed with this algorithm allow a very satisfactory extraction of the
vector-meson decay constants, with an accuracy that is certainly
competitive to that found using lattice QCD.

\section{Operator product expansion and choice of scheme for
heavy-quark masses} We start with the OPE for the correlation
function (\ref{1.1}). We may use the perturbative spectral density
$\rho_{\rm pert}(s,M)$ of \cite{chetyrkin} in terms of the pole
mass~of the heavy quark. An alternative option is to reorganize
the perturbative expansion~in terms of the running $\overline{\rm
MS}$ mass; the relevant analytic expressions are given in
\cite{kh}, see also the discussion in the Appendix.

Figure~\ref{Plot:1} illustrates the sum-rule estimates for
$f_{D^*}$ arising from (\ref{sr}) for these two choices of the
$c$-quark mass: the pole mass $M_c$ and the running $\overline{\rm
MS}$ mass $\overline{m}_c(\mu)$. The numerical OPE-parameter
values entering this game read~\cite{jamin,lms_plb2011,PDG,FLAG}
\begin{eqnarray}
\label{Table:1} 
&&\overline{m}_c(\overline{m}_c)=(1.275\pm 0.025)\;{\rm GeV},\quad 
\overline{m}(2\;{\rm GeV})=(3.42\pm 0.09)\;{\rm MeV},\quad 
\overline{m}_s(2\;{\rm GeV})=(93.8\pm 2.4)\;{\rm MeV},
\nonumber\\
&&\alpha_{\rm s}(M_Z)=0.1184\pm 0.0020,\\ 
&&\langle\bar
qq\rangle(2\;{\rm GeV})=-((267\pm 17)\;{\rm MeV})^3,\quad
\langle\bar ss\rangle(2\;{\rm GeV})/\langle\bar qq\rangle(2\;{\rm
GeV})=0.8\pm 0.3, \quad\left\langle\frac{\alpha_{\rm s}}{\pi}GG\right\rangle=(0.024\pm 0.012)\;{\rm GeV}^4. 
\nonumber
\end{eqnarray}
The pole mass, recomputed from the $O(\alpha_{\rm s}^2)$
relation between $\overline{m}_c$ and $M_c$ \cite{melnikov}, reads $M_c=1.699\;{\rm GeV}$. 
The sum-rule estimates shown in Fig.~\ref{Plot:1} are obtained for a $\tau$-independent
effective threshold $s_{\rm eff}$. Its values, different for
pole-mass OPE and $\overline{\rm MS}$-mass OPE, are found by
requiring maximal stability of the extracted decay constant in the
chosen Borel window (as detailed in Sect.~3). Let us emphasize that, for the moment, a \emph{constant\/}
effective threshold and the stability criterion for determining
its numerical value are adopted only for illustration: As we have
demonstrated in many examples \cite{lms_1}, using a constant
effective threshold provides rather inaccurate estimates for the
decay constant and does not allow one to probe the systematic
error of this extraction. 

Nevertheless, the results of Fig.~\ref{Plot:1} illustrate some of the essential features of the extraction procedures.  
First, using the pole-mass OPE, one observes no hierarchy of the perturbative contributions to the dual correlator --
the $O(1)$, $O(\alpha_s)$, and $O(\alpha_s^2)$ contributions have the same size. Obviously, there is no 
reason to expect the unknown higher-order perturbative corrections to be small;  
the pole-mass OPE truncated at order $O(\alpha_s^2)$ and the corresponding ground-state parameters suffer from large uncertainties. 
On the other hand, reorganizing the perturbative 
expansion in terms of the $\overline{\rm MS}$ mass of the heavy quark leads to a clear hierarchy and allows a reliable extraction 
of the ground-state parameters. This is precisely the same feature that has been observed for~the pseudoscalar correlator. 

Second, there is a huge numerical difference between the decay constants obtained using the pole-mass OPE and the running-mass OPE 
if one compares calculations obtained for the values of $\overline{m}_c(\overline{m}_c)$ and its pole-mass $O(\alpha_s^2)$ 
counterpartner given above. 
However, comparing the results of the truncated pole-mass and running-mass OPE requires some caution, 
as the perturbative expansion of the pole mass in terms of the running 
mass displays its asymptotic nature already at lowest orders \cite{melnikov}:  
$M_c=\overline{m}_c(\overline{m}_c)(1+1.33\, a + 10.32\, a^2 +116.50, a^3)$, with $a=\alpha_s(\overline{m}_c)/\pi=0.126\pm 0.002$. 
Assigning the uncertainty of the pole-mass value that corresponds to a specific running-mass value as, e.g., the size of the last 
included term in the perturbative relation, in our case of the $O(\alpha_s^2)$ term, amounts to a 15\% uncertainty in $M_c$. 
Due to a large sensitivity of the extracted decay constant to the precise value of the charm-quark mass, the uncertainty of 15\% 
in $M_c$ leads to a 100\% uncertainty in the dual pole-mass correlator. With such an uncertainty, the results obtained from the 
pole-mass and the running-mass OPE in Fig.~\ref{Plot:1} are compatible with each other, but suggest that the accuracy of the 
$O(\alpha_s^2)$-truncated pole-mass OPE is rather bad. 

We therefore make use of the OPE in terms of the running $\overline{\rm MS}$ mass for the analysis of $f_V$.
Accordingly, henceforth the quark masses $m_Q$ and $m$, and the strong coupling $\alpha_{\rm s}$ denote the $\overline{\rm MS}$
running quantities.

\begin{figure}[t]
\begin{tabular}{cc}
\includegraphics[width=8.cm]{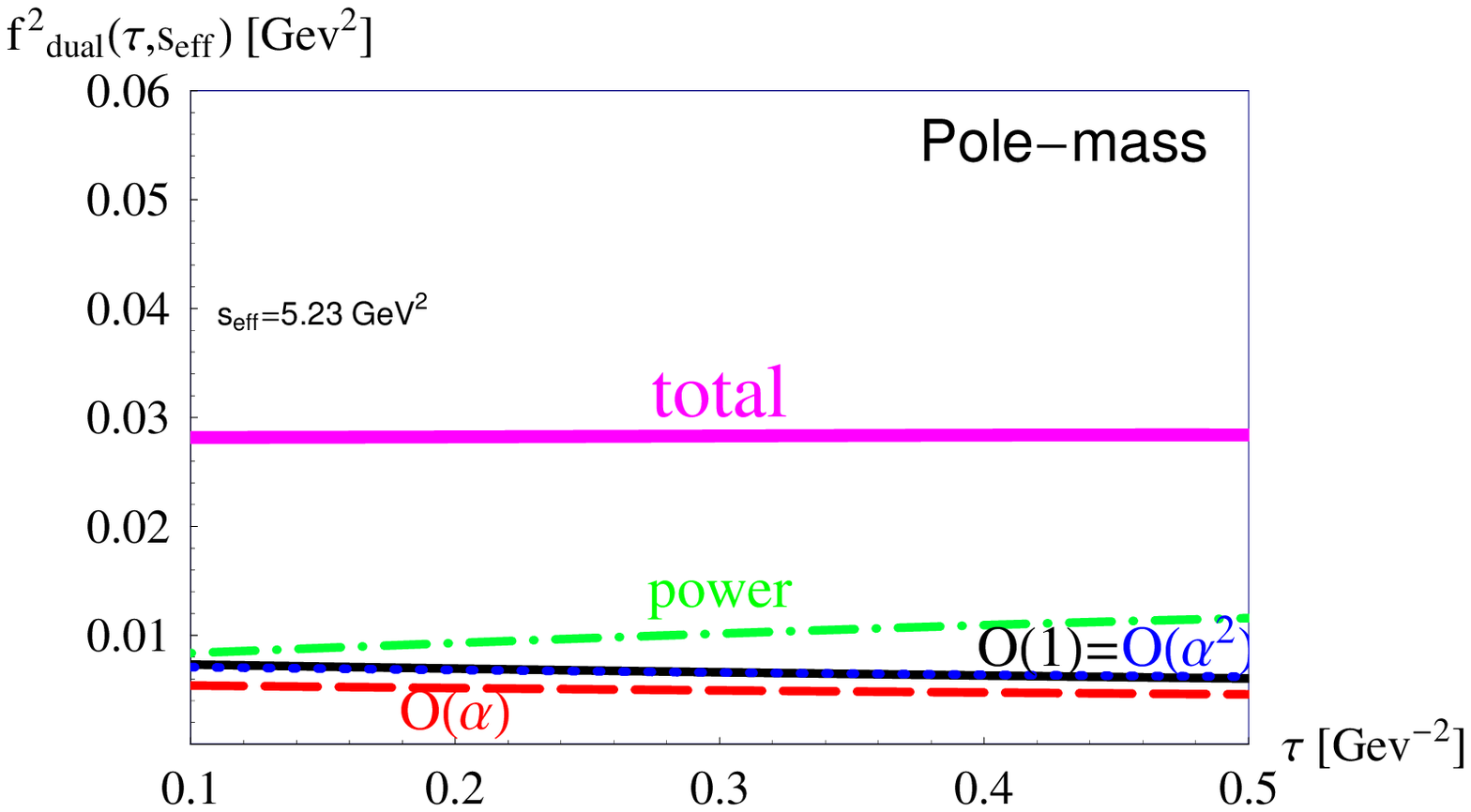}
&
\includegraphics[width=8.cm]{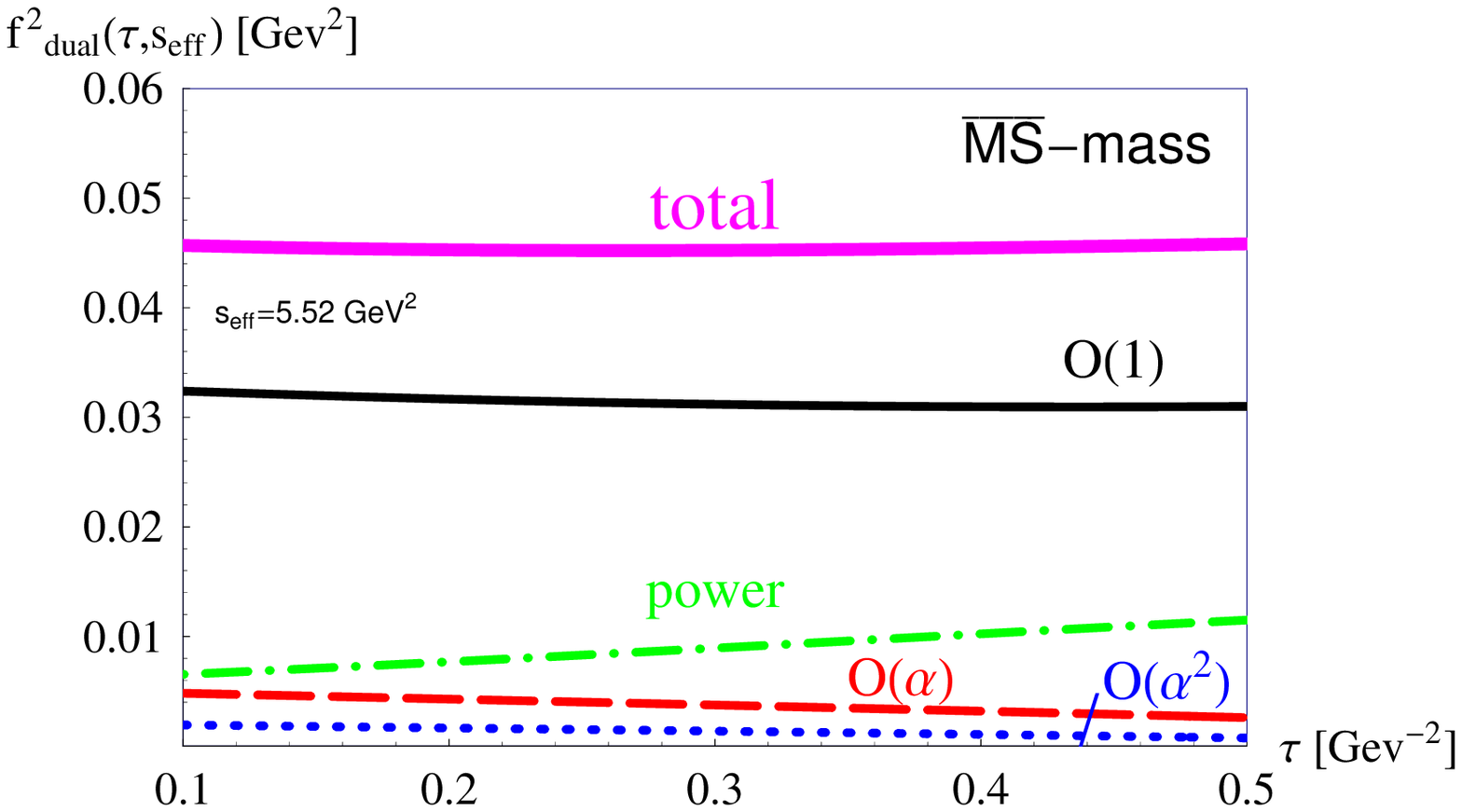}
\\
\end{tabular}
\caption{\label{Plot:1} QCD sum-rule estimates for 
$f_{D^*}$ extracted by expressing the OPE in terms
of the $c$-quark pole mass (left) or $\overline{\rm MS}$ mass
(right). The pole mass $M_c=1.699$ GeV, used in the left plot, has
been recalculated by the $O(\alpha_s^2)$ relation (\ref{a2}) from the running
$\overline{\rm MS}$ mass $m_c(m_c)=1.279$ GeV, used in the right
plot. For each case separately, a constant effective continuum
threshold~$s_{\rm eff}$ is determined by requiring ``maximal
stability'' of the obtained decay constant in the Borel window
$0.1\le \tau({\rm GeV}^{-2})\le 0.5$. As a result, $s_{\rm eff}$
turns out to be different for~the two schemes: $s_{\rm eff}=5.23$
GeV$^2$ for the pole-mass scheme (left), $s_{\rm eff}=5.52$
GeV$^2$ for the $\overline{\rm MS}$ scheme (right). Bold lines ---
total results, solid lines (black) --- $O(1)$ contributions;
dashed lines (red) --- $O(\alpha_{\rm s})$ contributions; dotted
lines (blue) --- $O(\alpha_{\rm s}^2)$ contributions; dot-dashed
lines (green) --- power contributions.}
\end{figure}

\section{Extraction of the decay constants}
In order to extract the decay constants from our QCD sum rule, we
first have to fix the working $\tau$-window where the OPE provides
a sufficiently accurate description~of the exact correlator (i.e.,
all higher-order radiative and power corrections are under
control) and the ground state gives a ``sizable'' contribution to
the correlator. We shall adopt~the window fixed in our previous
analysis of the decay constants of the $D$ and $D_s$ mesons
\cite{lms_plb2011}.

Next, we must fix the effective continuum threshold $s_{\rm
eff}(\tau)$. The corresponding algorithm was developed and
verified in quantum-mechanical potential models
\cite{lms_new,lms_qcdvsqm} and proven to work successfully for the
decay constants of the heavy pseudoscalar mesons \cite{lms_fp}.

We define the {\em dual invariant mass\/} $M_{\rm dual}$ and the
{\em dual decay constant\/} $f_{\rm dual}$ by
\begin{eqnarray}
\label{mdual} M_{\rm dual}^2(\tau) \equiv -\frac{d}{d\tau}\log
\Pi_{\rm dual}(\tau, s_{\rm eff}(\tau)),\qquad
\label{fdual} f_{\rm dual}^2(\tau)\equiv M_V^{-2}
e^{M_V^2\tau}\Pi_{\rm dual}(\tau, s_{\rm eff}(\tau)).
\end{eqnarray}
For a properly constructed $\Pi_{\rm dual}(\tau, s_{\rm eff}(\tau))$,
the dual mass coincides with the actual ground-state
mass $M_V$. Therefore, any deviation of the dual mass from $M_V$
is an indication of the contamination of the dual correlator by
excited states.

For any trial functional form of the effective threshold, one
obtains a variational solution by minimizing the difference
between the dual mass (\ref{mdual}) and the actual (experimental)
mass in the Borel window. This variational solution provides the
decay constant then via (\ref{fdual}). We consider a set of
$\tau$-dependent Ans\"atze for the effective continuum
threshold,~viz.,
\begin{eqnarray}
\label{zeff} s^{(n)}_{\rm eff}(\tau)=
\sum\limits_{j=0}^{n}s_j^{(n)}\tau^{j},
\end{eqnarray}
and fix the parameters on the right-hand side of (\ref{zeff})
by minimizing
\begin{eqnarray}
\label{chisq} \chi^2 \equiv \frac{1}{N} \sum_{i = 1}^{N} \left[
M^2_{\rm dual}(\tau_i) - M_V^2 \right]^2
\end{eqnarray}
over the window. This gives us the coefficients $s_j^{(n)}$ of the
effective continuum threshold and thus eventually the decay
constant~$f_V$. Still, different Ans\"atze for $s_{\rm eff}(\tau)$
yield different predictions for the decay constant.

A detailed analysis of quantum-mechanical models for different
potentials indicated that it is sufficient to consider polynomials
up to third order: In this case, the band delimited by the results
obtained for linear, quadratic, and cubic Ans\"atze for $s_{\rm
eff}(\tau)$ contains the true value of the decay constant. Even
the good knowledge of the truncated OPE~does not allow to
determine the decay constant precisely, but it allows us to
provide the range of values containing the true value of this
decay constant. The width of this range may be then treated as a
{\it systematic error} related to a principally limited accuracy
of the method. Presently, we do not see other possibilities to
obtain a more reliable estimate for the systematic~error.
Noteworthy, considering a merely $\tau$-independent threshold does
not allow one to probe the accuracy of the obtained estimate for
$f_V$.

%
%

On top of the systematic error comes the {\it OPE-related error} of the decay
constant: the OPE parameters are known with some~errors, inducing
a corresponding error of $f_V$. This OPE-related (or statistical)
error is determined by averaging the results for the decay
constant assuming for the OPE parameters Gaussian distributions
with the central values and standard deviations quoted in
(\ref{Table:1}) and a flat distribution over the scale $\mu$ in the
range $1< \mu\;({\rm GeV}) <3$.


\subsection{\boldmath Decay constant of the $D^*$ meson}

\begin{figure}[t]
\begin{tabular}{ccc}
\includegraphics[width=5.75cm]{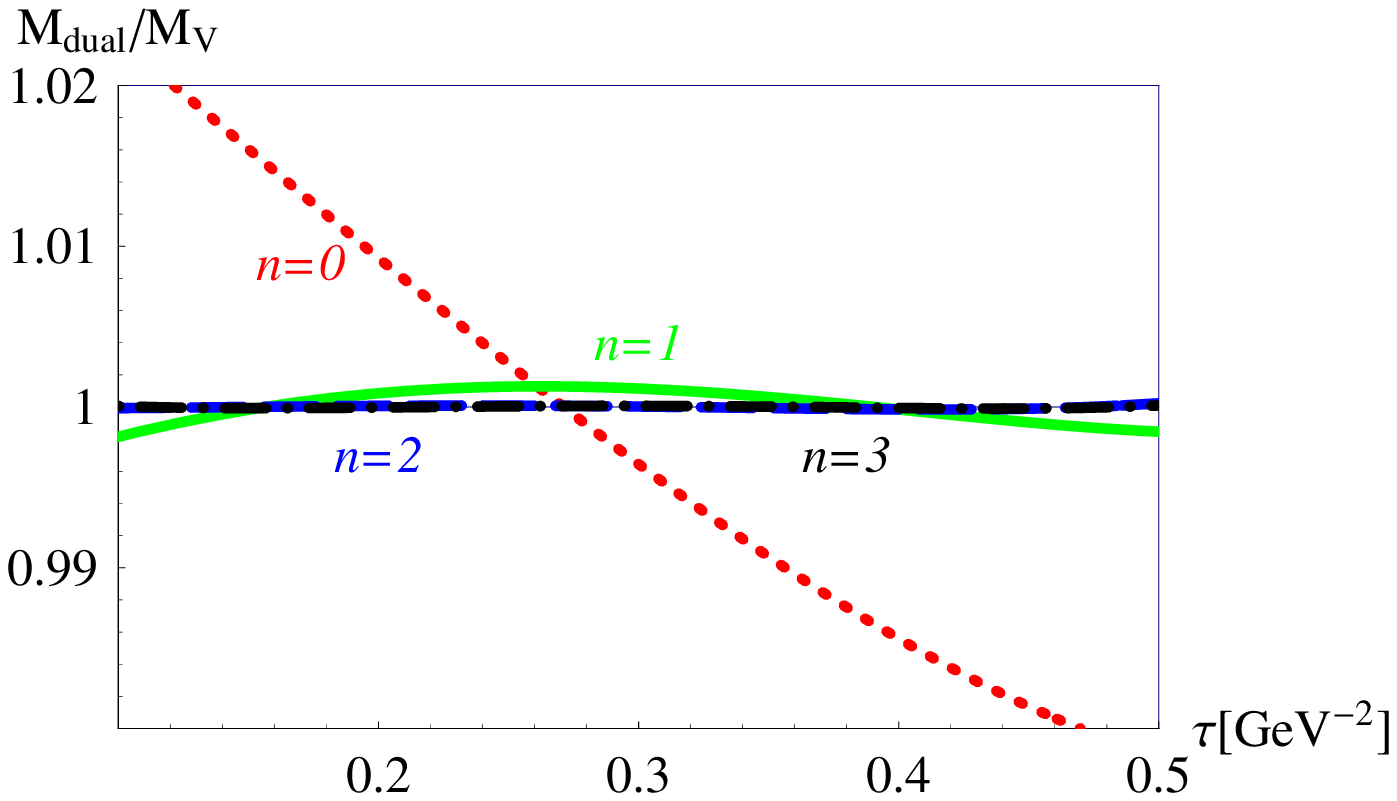}&
\includegraphics[width=5.75cm]{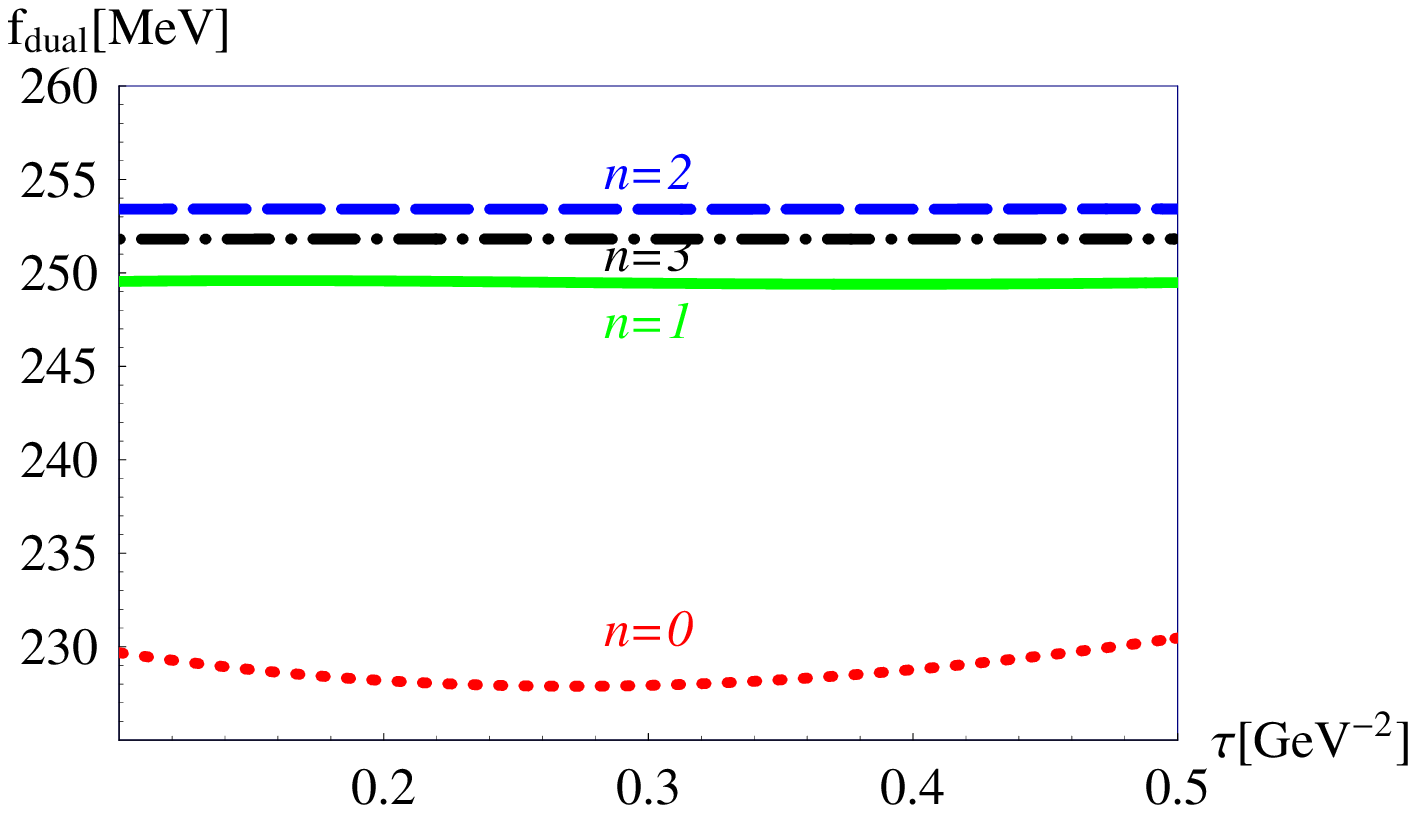}&
\includegraphics[width=5.75cm]{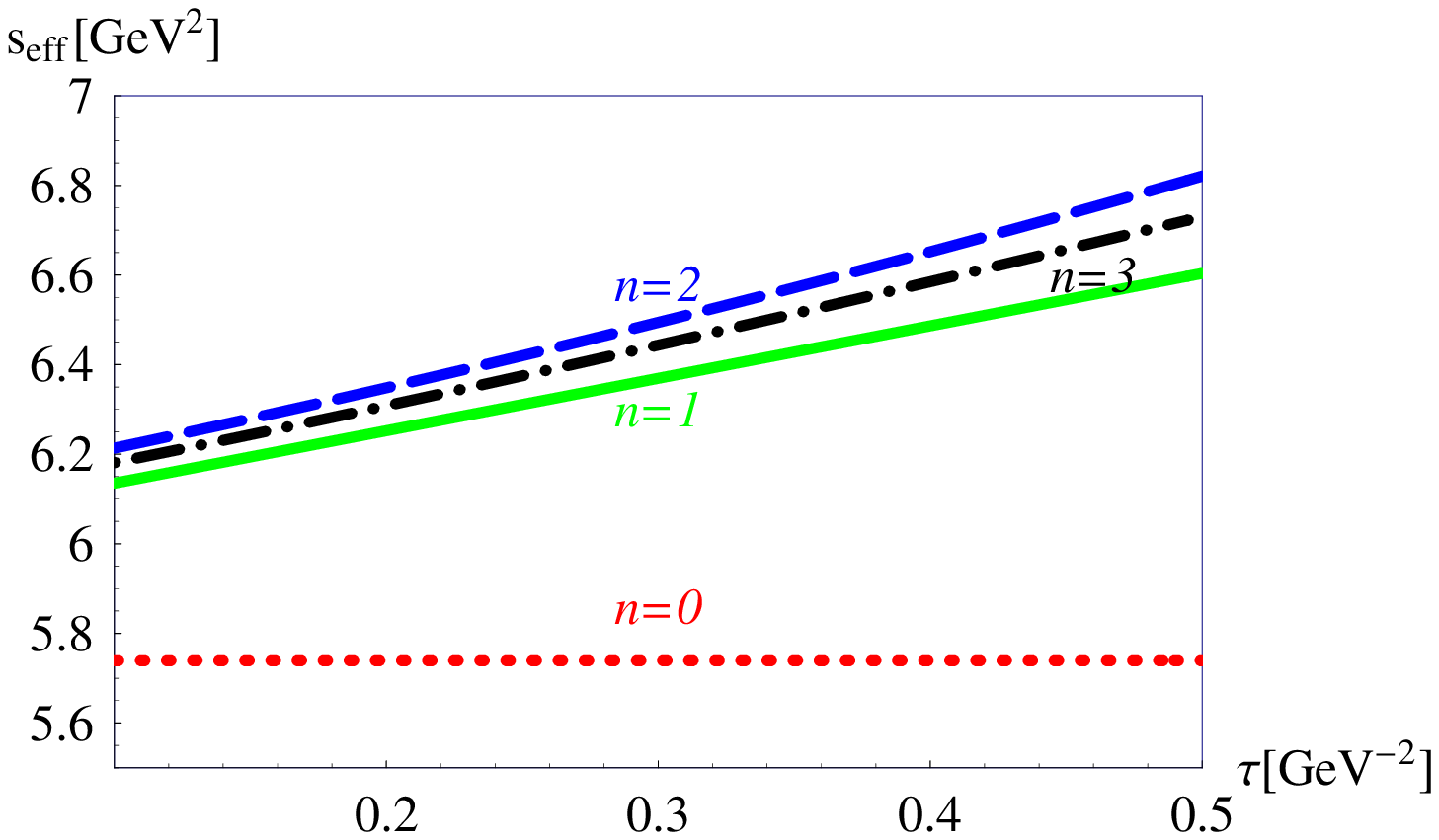}\\
(a) & (b) & (c)
\end{tabular}
\caption{\label{Plot:fD} Dependence on the Borel parameter $\tau$
of the dual mass (a) and the dual decay constant (b) of the $D^*$
meson, obtained~by employing different Ans\"atze (\ref{zeff}) for
the effective continuum threshold $s_{\rm eff}(\tau)$ and fixing
all thresholds according to (\ref{chisq});~the results are
presented for central values of all OPE parameters and for an
average scale $\mu=\mu^*=1.84$ GeV, where the average scale
$\mu^*$ is defined by (\ref{muaverage}). (c) Our $\tau$-dependent
effective thresholds obtained by the fitting procedure as
explained in the text. The integer $n=0,1,2,3$ is the degree of
the polynomial in our Ansatz (\ref{zeff}) for $s_{\rm eff}(\tau)$:
dotted lines (red) --- $n=0$; solid lines (green) --- $n=1$;
dashed lines (blue) --- $n=2$; dot-dashed lines (black) ---
$n=3$.}
\end{figure}

Following \cite{lms_plb2011}, we choose for the $\tau$-window for
the charmed mesons the interval
$\tau=(0.1$--$0.5)\;\mbox{GeV}^{-2}$. Figure~\ref{Plot:fD}~shows
the application of our procedure for fixing the effective
continuum threshold and extracting the resulting $f_{D^*}$.
As must be obvious from Fig.~\ref{Plot:fD}a, using a constant
threshold leads to a contamination of the dual correlator by
excited~states (at a percent level in the dual mass) while this
contamination is strongly reduced for $n>0$. The results for the
decay constant in Fig.~\ref{Plot:fD}b corresponding to $n>0$ are
nicely grouped together, whereas the $n=0$ prediction
lies~$\approx 30$~MeV below. Interestingly, the effect visible at
only a 1--2\% level in the dual mass in Fig.~\ref{Plot:fD}a
manifests itself at a 10\%~level in the decay constant in
Fig.~\ref{Plot:fD}b. Consequently, the results obtained for $n>0$,
less contaminated by excited states, constitute a significant
improvement with respect to the results obtained for a constant
threshold, i.e., $n=0$. Allowing the effective threshold to depend
on $\tau$
brings the QCD sum-rule results into agreement with the recent lattice
finding $f_{D^*}=(278\pm13\pm10)\;\mbox{MeV}$~\cite{lattice}.

The dependence of the extracted $f_{D^*}$ on both $c$-quark mass
$m_c\equiv {\overline m}_c({\overline m_c})$ and quark condensate
$\langle \bar qq\rangle\equiv \langle \bar qq(2\;{\rm
GeV})\rangle$ at the average scale $\mu^*=1.84$ GeV (see
(\ref{mustarDstar}) below) may be parameterized as
\begin{equation}
f_{D^*}^{\rm dual}(\mu=\mu^*,m_c,\langle \bar qq\rangle) =
\left[252.2 - 
10\left(\frac{m_c-\mbox{1.275\;GeV}}{\mbox{0.025\;GeV}}\right) + 6
\left(\frac{|\langle \bar
qq\rangle|^{1/3}-\mbox{0.267\;GeV}}{\mbox{0.01\;GeV}}\right) \pm
4_{\rm (syst)} \right] \mbox{MeV}.
\end{equation}

\begin{figure}[ht]
\begin{tabular}{cc}
\includegraphics[width=7cm]{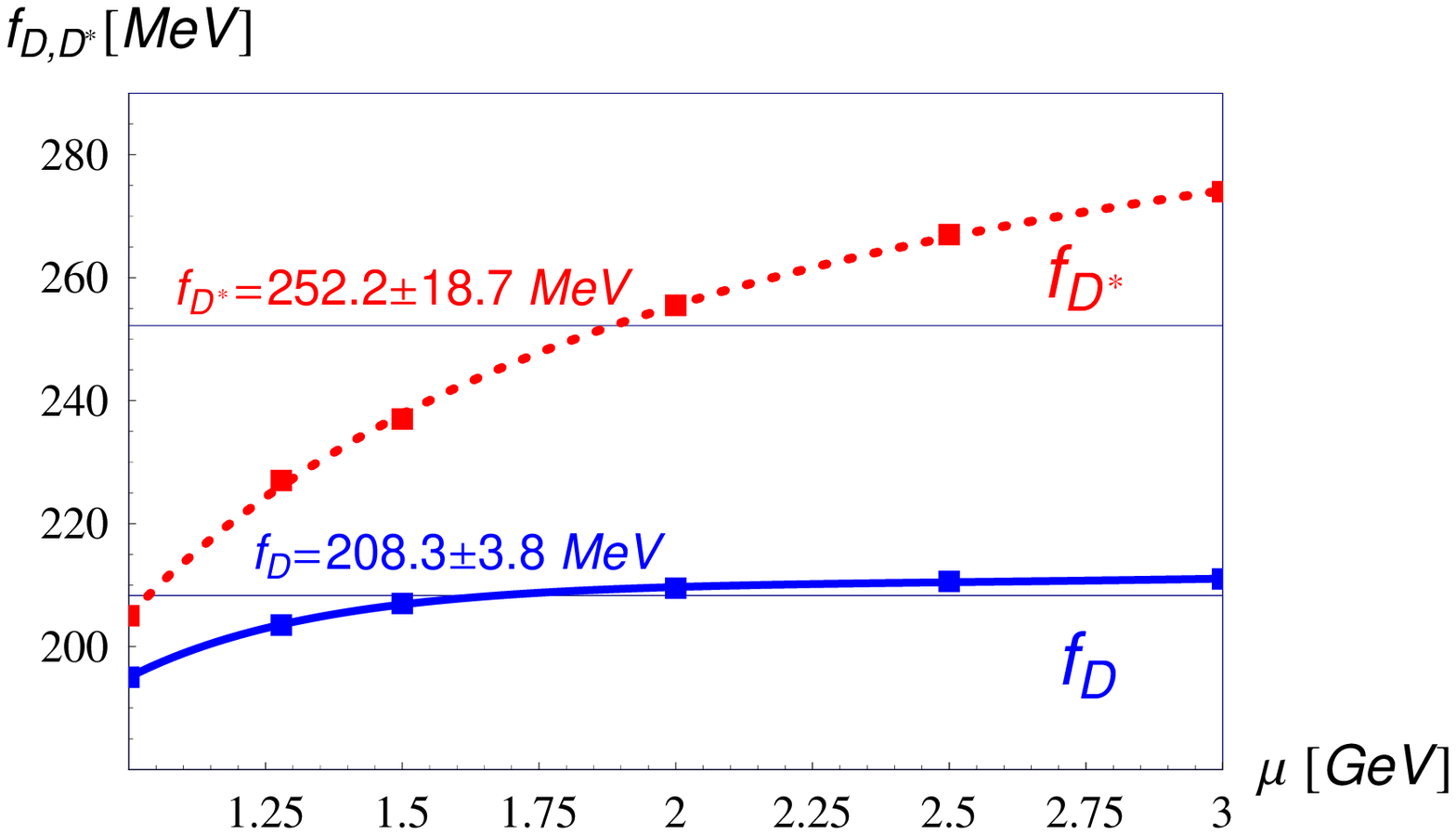}&
\includegraphics[width=7cm]{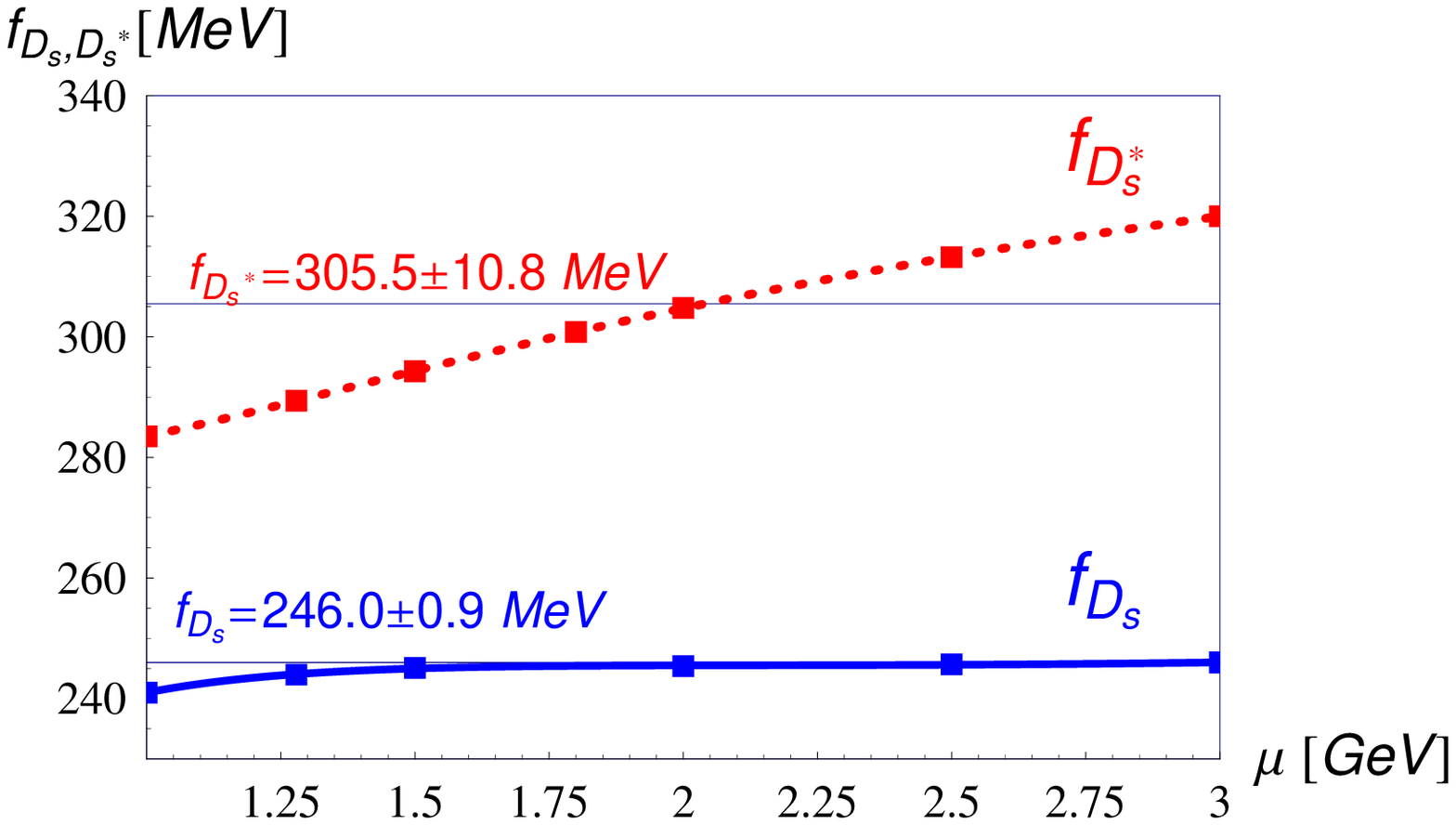}\\
(a) & (b)
\end{tabular}
\caption{\label{Plot:fV_vs_mu}
Dependence on $\mu$ of the dual
decay constants: (a) $f^{\rm dual}_D(\mu)$ and $f^{\rm
dual}_{D^*}(\mu)$, (b) $f^{\rm dual}_{D_s}(\mu)$ and $f^{\rm
dual}_{D^*_s}(\mu)$. The depicted results are obtained as follows:
for a fixed value of $\mu$, central values of the OPE parameters
in (\ref{Table:1}) and a Borel parameter $\tau$ within the window
$0.1 <\tau\;({\rm GeV}^{-2}) <0.5$, we determine the effective
thresholds by our procedure; the presented dual decay
constant~then is the average of the band formed by the linear,
quadratic, and cubic Ans\"atze for the effective threshold.
Clearly, the effective thresholds turn out to depend on the scale
$\mu$. Dotted lines (red) --- vector mesons; solid lines (blue)
--- pseudoscalar mesons.}
\end{figure}

The extracted value of $f_{D^*}$ turns out to be very sensitive to
the choice of the renormalization scale $\mu$. Recall once more
that this dependence is unphysical and induced by the truncation
of the perturbation series. The $\mu$ dependence of $f_{D^*}$ for
the central values of the other OPE parameters is depicted in
Fig.~\ref{Plot:fV_vs_mu}a. For each $\mu$, the value of $f_{D^*}$
(and $f_{D}$) corresponds to the average of the interval formed by
the results obtained from the linear, quadratic, and
cubic~Ans\"atze for the effective continuum threshold. It should
be noted that the dependence of $f_{D^*}$ on $\mu$ is clearly
nonlinear.
The obtained results may be well interpolated by the following
simple formula:
\begin{equation}
\label{fDsmu}
f_{D^*}^{\rm dual}(\mu) = 252.2 \mbox{ MeV}
\left[1+0.233 \log(\mu/\mu^*)-0.096 \log^2(\mu/\mu^*)+0.17
\log^3(\mu/\mu^*)\right], \qquad \mu^*=1.84\mbox{ GeV}.
\end{equation}
Here, $\mu^*$ is the average scale defined in the standard way:
\begin{equation}
\label{muaverage} \langle f_V^{\rm dual}(\mu)\rangle=f_V^{\rm
dual}(\mu^*),
\end{equation}
assuming a flat probability distribution for $\mu$ in the range
$1<\mu\;({\rm GeV})<3$. The corresponding standard deviation~of
$f_{D^*}$ is $18.7$ MeV. For comparison, we also provide the $\mu$
dependence and the average scale $\mu^*$ for $f_D$ from
\cite{lms_plb2011}:
\begin{equation}
\label{mustarDstar}
f_{D}^{\rm dual}(\mu) = 208.3 \mbox{ MeV}\left[1+0.06
\log(\mu/\mu^*)-0.11 \log^2(\mu/\mu^*)+0.08
\log^3(\mu/\mu^*)\right], \qquad \mu^*=1.62\mbox{ GeV}.
\end{equation}
Obviously, the $\mu$ dependence of the pseudoscalar correlator is
much weaker. This effect has the following origin: both the
truncated perturbative dual correlator $\Pi^{\rm dual}_{\rm
pert}(s_{\rm eff},\tau,\mu)$ and the truncated $\Pi_{\rm
power}(\tau,\mu)$ exhibit a rather pronounced $\mu$ dependence.
For the pseudoscalar correlator, these $\mu$ dependencies to a
large extent cancel each other, whereas~for~the vector correlator
the cancellation does not occur.

\begin{figure}[t]
\begin{tabular}{ccc}
\includegraphics[width=5.7cm]{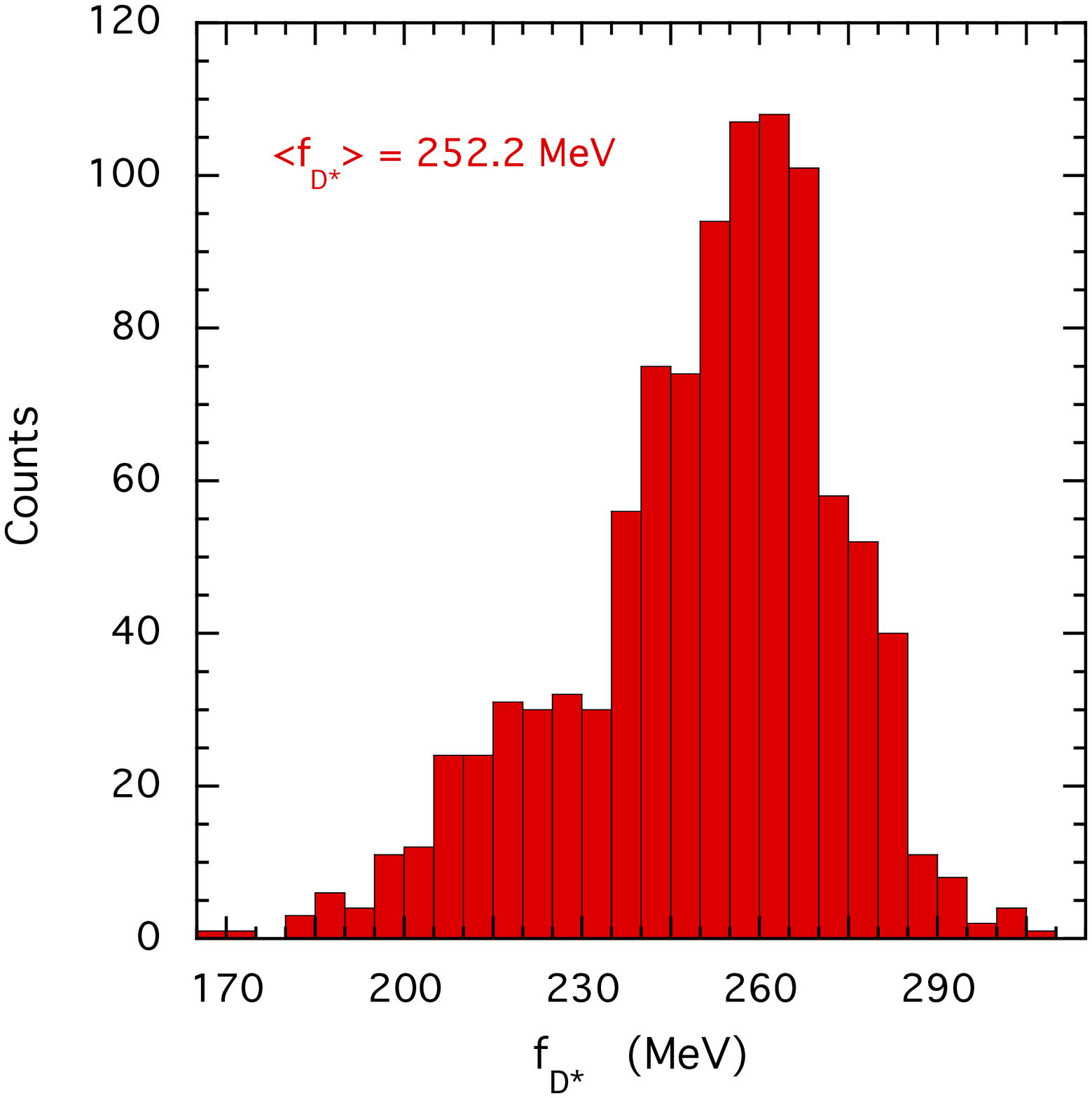}&
\includegraphics[width=5.7cm]{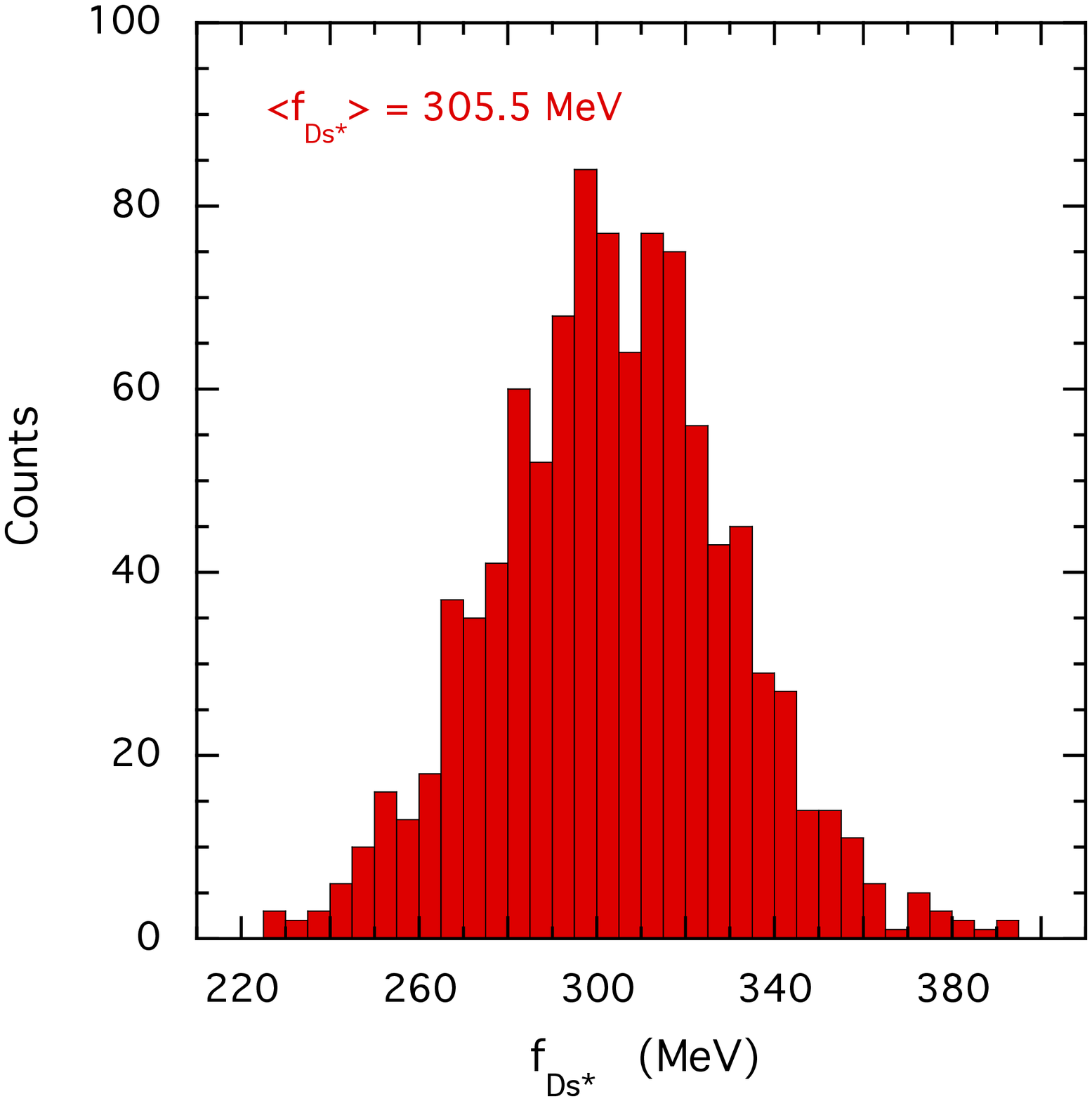}
\end{tabular}
\caption{\label{Plot:bootstrap} Distribution of the decay
constants $f_{D^*}$ (a) and $f_{D_s^*}$ (b) obtained by 1000
bootstrap events.}
\end{figure}

Assuming Gaussian distributions for all the OPE parameters
collected in (\ref{Table:1}) and a flat $\mu$ distribution in the
range $1 < \mu\;(\rm GeV) <3$, we obtain the distribution of
$f_{D^*}$ depicted in Fig.~\ref{Plot:bootstrap}. The $f_{D^*}$
distribution is clearly not Gaussian, which is due to the
nonlinear $\mu$ dependence of $f_{D^*}$ shown in
Fig.~\ref{Plot:fV_vs_mu}. For the average and the standard
deviation of~the $D^*$-meson decay constant we obtain
\begin{eqnarray}
\label{fDv_constant} 
f_{D^*} = \left(252.2\pm 22.3_{\rm (OPE)}\pm
4_{\rm (syst)}\right) \mbox{MeV}.
\end{eqnarray}
The OPE uncertainty is composed as follows: 18.7 MeV are due to
the variation of the scale $\mu$, 10 MeV arise from the error in
$m_c\equiv {\overline m}_c({\overline m_c})$, 2 MeV from
$\alpha_{\rm s}(M_Z)$, 6 MeV from the quark condensate, and 3 MeV
from the gluon~condensate. Higher condensates contribute less than
1 MeV to this error.

Combining our above results with those for $f_D$ from our earlier
analysis \cite{lms_plb2011}, we obtain
\begin{eqnarray}
\label{fDvs/fDs}
f_{D^*}/f_D = 1.221\pm 0.080_{\rm (OPE)}\pm 0.008_{\rm (syst)}.
\end{eqnarray}
The OPE uncertainty of this ratio is fully dominated by the impact
of the $\mu$ dependence.


\subsection{\boldmath Decay constant of $D^*_s$ meson}
\begin{figure}[b]
\begin{tabular}{ccc}
\includegraphics[width=5.75cm]{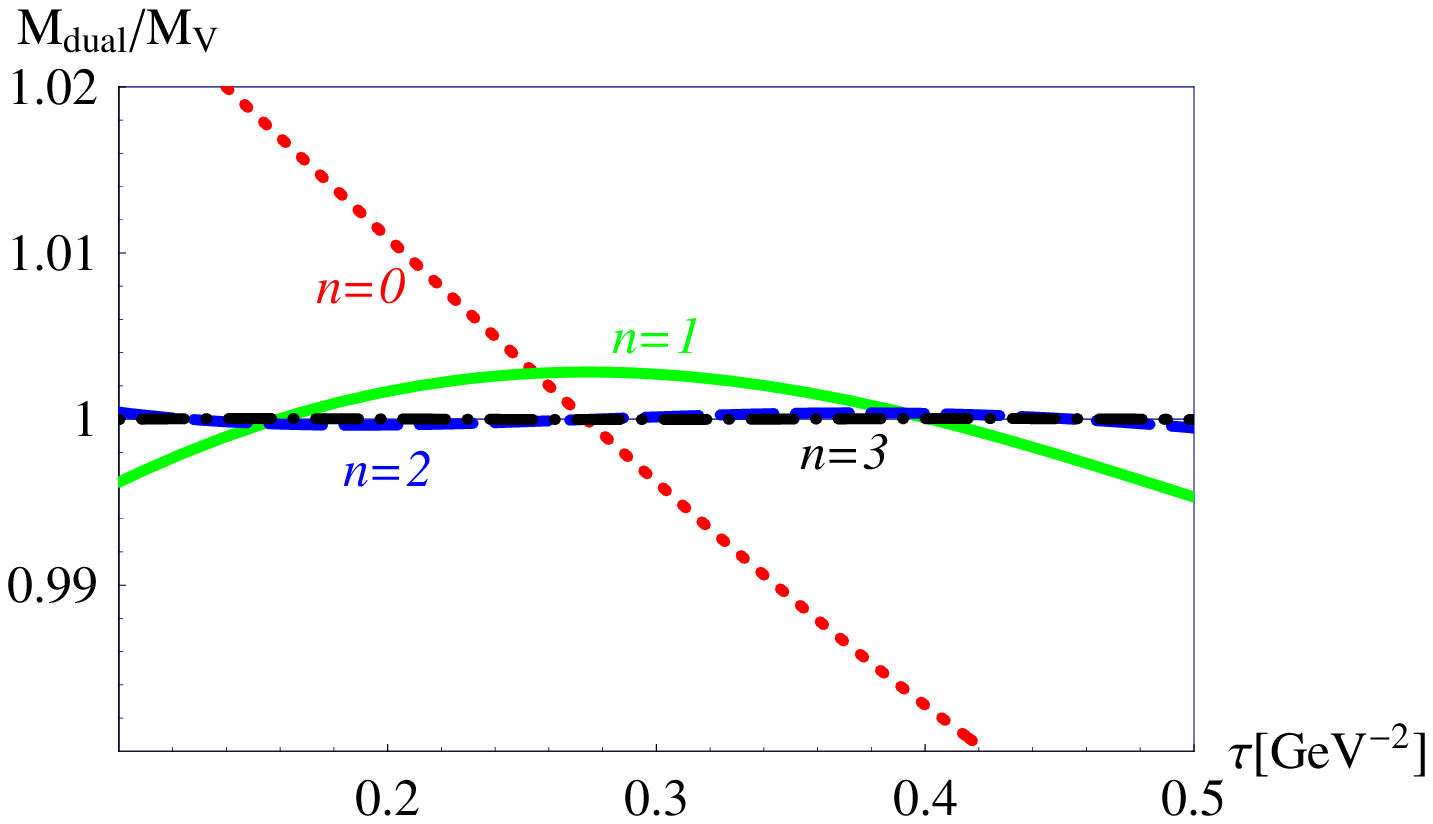}&
\includegraphics[width=5.75cm]{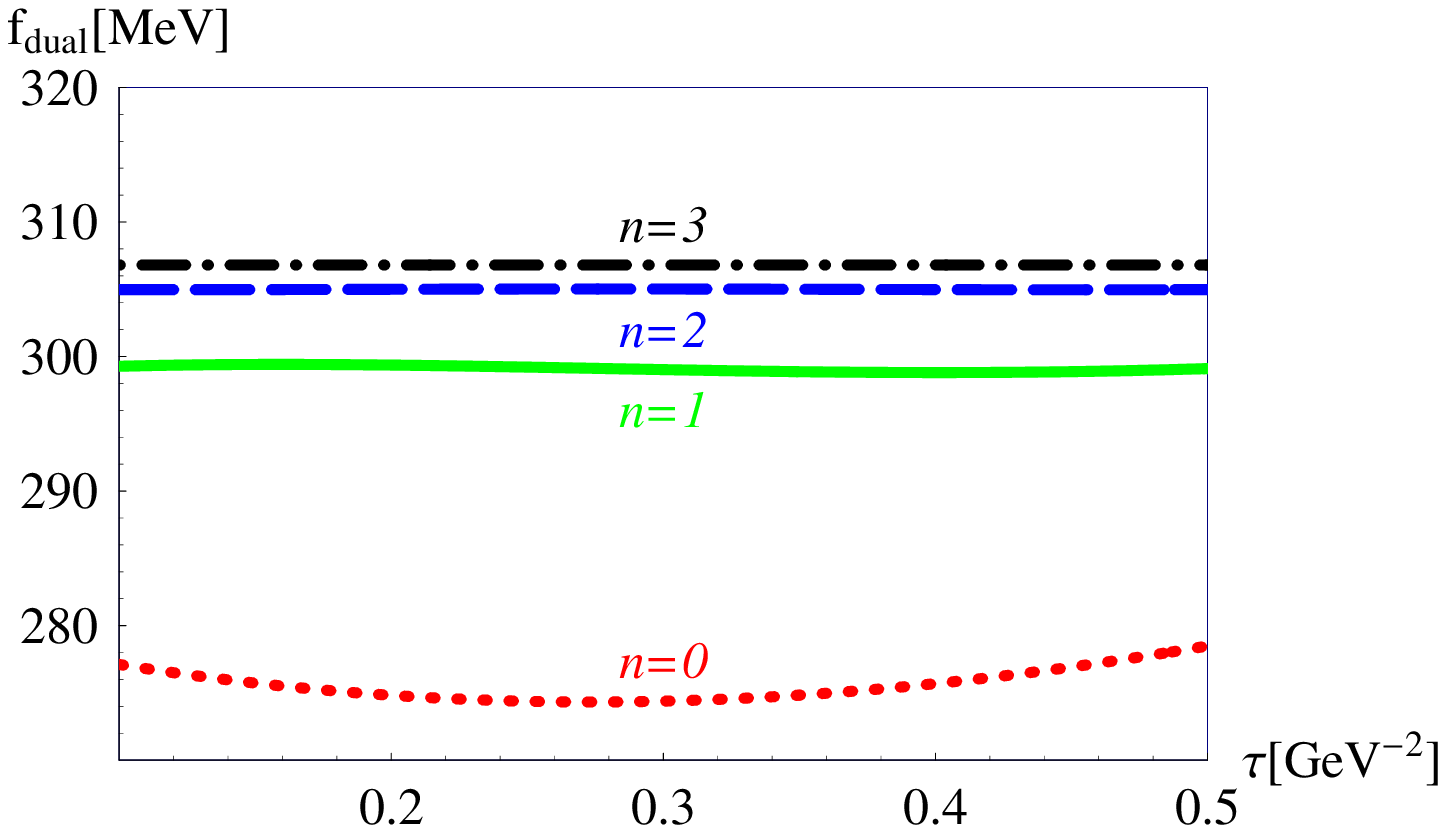}&
\includegraphics[width=5.75cm]{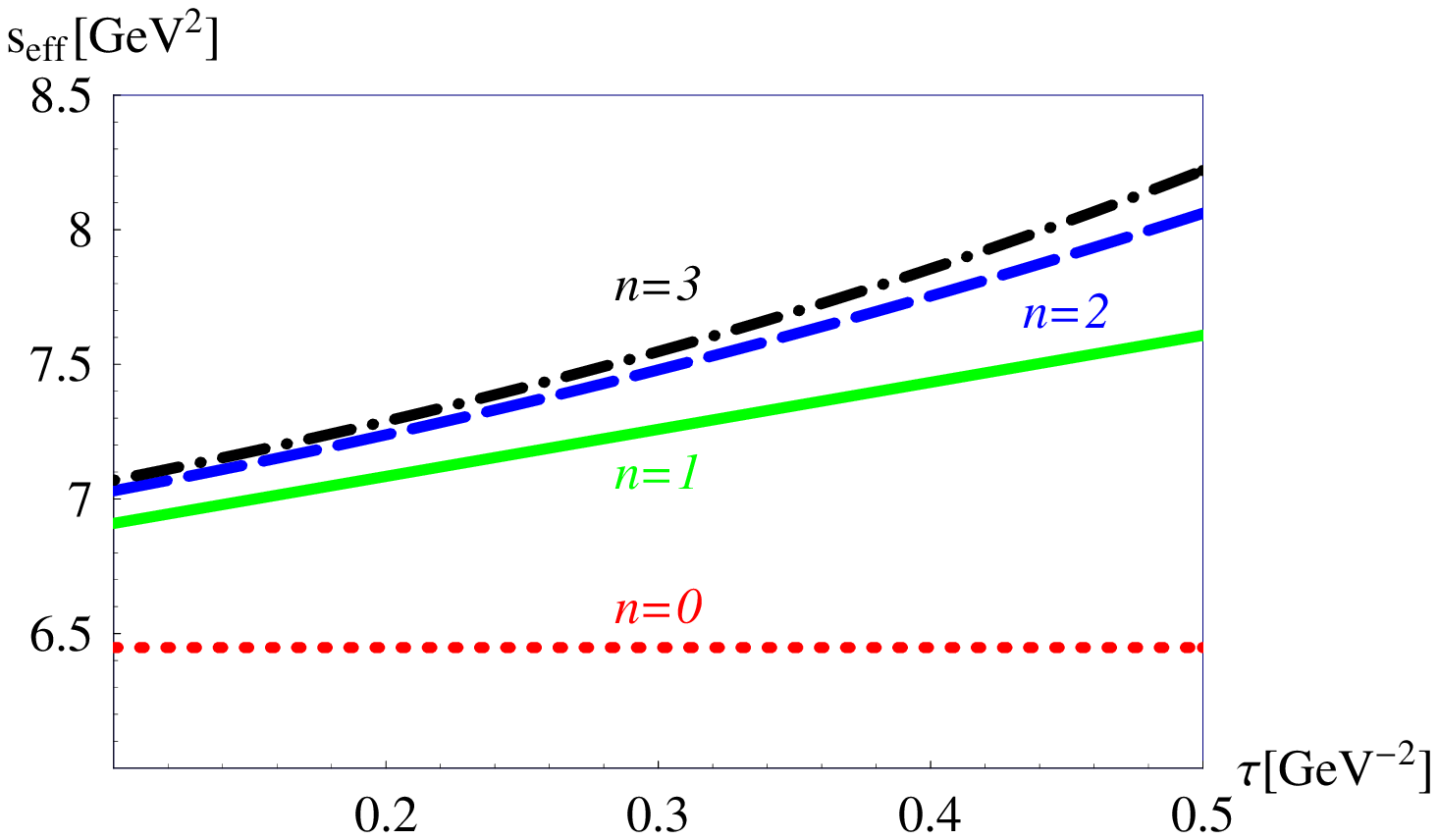}\\
(a) & (b) & (c)
\end{tabular}
\caption{\label{Plot:fDs} Same as Fig.~\ref{Plot:fD} but for the
$D^*_s$ meson, at the average renormalization scale appropriate
for the $D^*_s$ meson: $\mu^*=1.94\;\mbox{MeV}$.}
\end{figure}

For the $D^*_s$, we take the same Borel-parameter window as for
$D^*$: $\tau = (0.1$--$0.5)\;\mbox{GeV}^{-2}$.
Figure~\ref{Plot:fDs} provides the details of our extraction
procedure.
Our results for the $D^*_s$-meson decay constant may be summarized
as [$m_s\equiv\overline{m}_s(2\;\mbox{MeV})$]
\begin{align}
f_{D^*_s}^{\rm dual}(\mu=\mu^*,m_c,m_s,\langle \bar ss\rangle)
=&\left[305.5
-12.4\left(\frac{m_c-\mbox{1.275\;GeV}}{\mbox{0.025\;GeV}}\right)
+1.7\left(\frac{m_s-\mbox{0.1\;GeV}}{\mbox{0.004\;GeV}}\right)\right.
\nonumber\\&\left. +\;3.9 \left(\frac{|\langle \bar
ss\rangle|^{1/3}-\mbox{0.248\;GeV}}{\mbox{0.01\;GeV}}\right) \pm
5_{\rm (syst)} \right] \mbox{MeV}.
\end{align}

Similarly to $f_{D^*}$, also the extracted decay constant of
$D^*_s$ exhibits a rather strong and almost linear $\mu$
dependence (see Fig.~\ref{Plot:fV_vs_mu}b) which, for average
values of the other OPE parameters, may be parameterized as
\begin{equation}
\label{fDvsmu} 
f_{D_s^*}^{\rm dual}(\mu) = 305.5 \mbox{ MeV}
\left[1+0.124 \log(\mu/\mu^*)+0.014 \log^2(\mu/\mu^*)-0.034
\log^3(\mu/\mu^*)\right], \qquad \mu^*=1.94\mbox{ GeV}.
\end{equation}
For comparison, the $\mu$ dependence and the average scale $\mu^*$
for $f_{D_s}$ from \cite{lms_plb2011} is also given:
\begin{equation}
f_{D_s}^{\rm dual}(\mu) = 246.0 \mbox{ MeV}\left[1+0.01
\log(\mu/\mu^*)-0.03 \log^2(\mu/\mu^*)+0.04
\log^3(\mu/\mu^*)\right], \qquad \mu^*=1.52\mbox{ GeV}.
\end{equation}
Notice that $f_{D_s}$ is extremely stable with respect to $\mu$.
This is an effect of an almost precise cancellation between
the~$\mu$ dependencies of the dual perturbative and the condensate
contributions.

Again, for Gaussian distributions of all OPE parameters and a flat
distribution in $\mu$ in the range $1 < \mu\;(\mbox{GeV}) <3$, we
find a nearly Gaussian distribution of $f_{D_s^*}$ in
Fig.~\ref{Plot:bootstrap} which yields
\begin{eqnarray}
\label{fDvs_constant} 
f_{D^*_s} = \left(305.5\pm 26.8_{\rm(OPE)}\pm 5_{\rm (syst)}\right) \mbox{MeV}.
\end{eqnarray}
The composition of the OPE error reads: 10.8 MeV are due to the
variation of the scale $\mu$, 19.5 MeV are caused by the error of
strange-quark condensate, 12.5 MeV by the error of
$\overline{m}_c(\overline{m}_c)$, 6.4 MeV by the gluon condensate,
1.7 MeV by the strange-quark mass, and 1.4 MeV by $\alpha_{\rm
s}(M_Z)$. Higher condensates contribute 2 MeV to this uncertainty.
Our result (\ref{fDv_constant}) is in good agreement with
$f_{D_s^*} =(311\pm 9)$ MeV from lattice QCD \cite{lattice}.

Making use of our result for $f_{D_s}$ from \cite{lms_plb2011}, we
obtain, for the ratio of the vector and the pseudoscalar decay
constants,
\begin{eqnarray}
\label{fDv/fD}
f_{D_s^*}/f_{D_s} = 1.241\pm 0.057_{\rm (OPE)}\pm 0.007_{\rm (syst)}.
\end{eqnarray}
The OPE uncertainty in this ratio is dominated by the errors
arising from the $\mu$ dependence (0.043) and the gluon condensate
(0.026).

Finally, for the ratio of the $D_s^*$ and $D^*$ decay constants,
we get
\begin{eqnarray}
\label{ratioDv}
f_{D^*_s}/f_{D^*} = 1.211\pm 0.061_{(\rm OPE)}\pm 0.007_{(\rm syst)}.
\end{eqnarray}
The error here arises mainly from the errors in the strange-quark
mass and the condensates ratio $\langle \bar s s\rangle/\langle
\bar q q\rangle=0.8\pm 0.3$. The value (\ref{ratioDv}) is slightly
larger than but not in disagreement with the lattice result
$f_{D^*_s}/f_{D^*} = 1.16\pm 0.02\pm 0.06$~\cite{lattice}.\footnote{For an analysis of the vector-meson
decay constants within the framework of quark models, we refer to \cite{faustov}.}

\section{Summary and conclusions}
Exploiting the tools offered by QCD sum rules, we analyzed in
great detail the decay constants of charmed vector mesons, paying
special attention to the involved uncertainties of the predicted
decay-constant values: the OPE error (related to the precision
with which the QCD parameters are known) and the systematic error
intrinsic to the sum-rule approach as a whole (reflecting the
limited accuracy of the extraction procedure). We thus gained
important insights:

\begin{itemize}
\item[(i)] As was already noted in the case of heavy
pseudoscalar mesons \cite{lms_plb2011}, also for the vector
correlator the perturbative expansion in terms of the heavy-quark
pole-mass does not seem to converge whereas reorganizing it in
terms of the corresponding running mass leads to a clear hierachy
of the perturbative contributions.

\item[(ii)] The dependence of the vector correlator, known at
three-loop accuracy, on the renormalization scale $\mu$ turns~out
to be sizeably stronger compared to the pseudoscalar correlator.
Respectively, the error related to the remaining scale dependence
of the vector-meson decay constant proves to be twice as large as
that for pseudoscalar-meson decay constant.

\item[(iii)] We allowed for a Borel-parameter-dependent effective
threshold for the decay-constant extractions. Obviously, such a
$\tau$-dependent effective threshold visibly improves the
stability of the dual mass in the Borel window. This means that
the dual correlator is much less contaminated by excited states
than the one inferred upon~confining oneself to $\tau$-independent
effective thresholds. We thus get, as our estimates for the
vector-meson decay constants,
\begin{align}
f_{D^*}&=\left(252.2 \pm 22.3_{\rm (OPE)}\pm 4_{\rm(syst)}\right)
\mbox{ MeV},
\\
f_{D^*_s}&=\left(305.5\pm 26.8_{\rm (OPE)}\pm 5_{\rm(syst)}\right) \mbox{ MeV},
\end{align}
and, for the various ratios of decay constants,
\begin{align}
f_{D^*_s}/f_{D^*}&= 1.211\pm 0.061_{(\rm OPE)}\pm 0.007_{(\rm syst)},\\
f_{D^*}/f_D&= 1.221\pm 0.080_{\rm (OPE)}\pm 0.008_{\rm (syst)},\\
f_{D^*_s}/f_{D_s}&= 1.241\pm 0.057_{\rm (OPE)}\pm 0.007_{\rm (syst)}.
\end{align}
The OPE uncertainties in the decay constants of $D^*$ and $D^*_s$
and in the above ratios are, to large extent, due~to the remaining
dependence on the renormalization scale $\mu$.

Our predictions agree well with those from lattice QCD,
$f_{D^*}=(278\pm 13\pm 10)\;\mbox{MeV}$ and 
$f_{D^*_s}=(311\pm 9)\,\mbox{MeV}$~\cite{lattice}. 

Our results are in agreement with the recent estimates presented in Ref.~\cite{kh}, which also make use of our 
idea of a $\tau$-dependent effective threshold. However, in our opinion, the estimates of \cite{kh} are not 
fully trustworthy: first, the OPE used in \cite{kh} contained errors which we correct (see (\ref{A3}) and (\ref{A4}));
second, the authors of \cite{kh} do not take properly into account the $\tau$-dependence of 
the effectve threshold when calculating the dual mass. 

We stress that our algorithm for fixing $\tau$-dependent effective
thresholds allows us to provide, in addition to the OPE errors,
also the systematic errors intrinsic to the QCD sum-rule
technique. Although not entirely rigorous in the mathematical
sense, our algorithm for obtaining the systematic errors has been
verified in several examples within quantum mechanics, and proved
to work well for decay constants of pseudoscalar mesons. The good
news is that the systematic uncertainty turns out to be small and
to be under control.

\item[(iv)] The $\tau$-dependent thresholds entail a visible shift
in the sum-rule predictions for the decay constants of charmed
vector mesons, increasing their numerical values by roughly 30 MeV
compared to the outcomes when sticking~to~a constant threshold
determined by the criterion of stability in the same Borel
window. 
\end{itemize}

\vspace{3ex}{\bf Acknowledgements.} 
D.M.\ was supported by a grant
for leading scientific schools 3042.2014.2 (Russia). 
S.S. thanks MIUR(Italy) for partial support under contract No. PRIN 2010-2011.

\appendix\section{OPE for the vector correlator}
The perturbative spectral densities have been calculated in
three-loop order in \cite{chetyrkin} for one massless and one
massive quark in terms of the pole mass $M$ of the latter:
\begin{eqnarray}
\label{R} \rho_{\rm
pert}(s,M)=\rho^{(0)}(s,M)+a(\mu)\rho_1(s,M)+a^2(\mu)\rho_2(s,M),
\qquad a(\mu)\equiv \frac{{\alpha}_{\rm s}(\mu)}{\pi}.
\end{eqnarray}
We reorganize this expansion in terms of the related running mass 
$m_Q\equiv \overline{m}_Q(\mu)$ (using the notations of \cite{jamin}):
\begin{eqnarray}
\label{a2}
M=\frac{m_Q}{1+a(\mu)r^{(1)}_m+a^2(\mu)r_m^{(2)}}.
\end{eqnarray}
The corresponding spectral densities and the expressions for the
power corrections were taken from the Appendix to \cite{kh},
except for Eqs.~(A3) and (A4) therein, for which we obtain
different results:
\begin{align}
\label{A3} 
\Delta_1\rho_T^{\rm(pert,NNLO)}(s)&=-\frac{3}{8\pi^2}s
\,\bm{z}\left[(3-7z^2){r^{(1)}_m}^2-2(1-z^2)r_m^{(2)}\right],\\
\label{A4} \Delta_2\rho_T^{\rm(pert,NNLO)}(s)&
=-\frac{1}{16\pi^2}C_F r^{(1)}_m s\Big[-12z(1-z^2)\left(2{\rm
Li}_2(z)+\log(z)\log(1-z)\right)\nonumber\\
&\hspace{19.2ex}-2z(9+6z-17z^2)\log(z)\nonumber\\&\hspace{19.2ex}
+2(1-z)(\bm{-4+5z+17z^2})\log(1-z)-z(1-z)(17+15z)\Big],\\
z&\equiv\frac{{m}^2_Q}{s}.\nonumber
\end{align}
These equations replace the corresponding equations from \cite{kh}.

\end{document}